# Scaling up to Problem Sizes: An Environmental Life Cycle Assessment of Quantum Computing

Sylvain Cordier [a], Karl Thibault [b], Marie-Luc Arpin [c], and Ben Amor [a]

a: Interdisciplinary Research Laboratory on Sustainable Engineering and Ecodesign (LIRIDE), Department of Civil and Building Engineering, Université de Sherbrooke, Sherbrooke, QC J1K 2R1, Canada;

b: Institut quantique, Université de Sherbrooke, Sherbrooke, QC J1K 2R1, Canada;

c: Department of management and human resources management, Université de Sherbrooke, Sherbrooke, QC J1K 2R1, Canada;

* Corresponding authors: sylvain.cordier@usherbrooke.ca, ben.amor@usherbrooke.ca, karl.thibault@usherbrooke.ca

## Abstract

With the demonstrated ability to perform calculations in seconds that would take classical supercomputers thousands of years, quantum computers namely hold the promise of radically advancing sustainable IT. However, quantum computers face challenges due to the inherent noise in physical qubits, necessitating error correction for reliable operation in solving industrial-scale problems, which will require more computation time, energy, and electronic components than initial laboratory-scale experiments. Yet, while researchers have modeled and analyzed the environmental impacts of classical computers using Life Cycle Assessment (LCA), the environmental performance of quantum computing remains unknown to date. This study contributes to filling this critical gap in two ways: (1) by establishing an environmental profile for quantum computers based on superconducting qubits; and (2) by comparing it to a functionally equivalent profile of a state-of-the-art supercomputer. With the comparison based on the problem size, the paper shows how the usage time can drive an environmental advantage for quantum computers under specific scaling conditions and quantum error correcting codes. The results emphasize that quantum error correction hardware has a substantial environmental impact due to the numerous electronic components needed to achieve 100 logical qubits. This paper can serve as a basis for designing more environmentally friendly quantum computers and for establishing their environmental profiles, as well as those of the human activities that will use them.

# 1. Introduction

Quantum computing technologies are emerging as a new paradigm in high-performance computing with the potential to solve previously unsolvable problems using classical methods. However, it is yet unclear if quantum computing will surpass classical computing in terms of speed, as classical computing methods keep narrowing the gap with quantum computing methods only days after each breakthrough. For example, in 2019 a 53-qubit computer performed a 200-second calculation that would have taken 10,000 years on the Summit supercomputer [1]. IBM later estimated it would take 2.5 days [2], Huang *et al*. suggested their simulator could do it in less than 20 days using a Summit-comparable cluster [3], and even later others reached 15 hours [4]. Another example is boson sampling with a photonic computer from Xanadu which took 36 micro seconds instead of 8,500 years on the Fugaku supercomputer [5]. This literature review on examples of classical method developments, focusing on simulations of state-of-the-art quantum computing experiments and "quantum-inspired" algorithms by classical methods, is an example of this "friendly competitive escalation" between the two communities [6]. Nonetheless, studies addressing the energy consumption show an advantage of quantum computer over classical supercomputers [7], [8].

Be that as it may, quantum technology must deal with inherent noise in the physical qubits, necessitating error correction to create logical qubits. Such system is called a "fault-tolerant quantum computer". Insofar as it is still possible to use quantum computers without quantum error correction (QEC), i.e. so-called noisy intermediate-scale quantum computation approaches and imperfect qubits, the problem size should require few qubits overall [9] and small circuit depths. Such non-fault-tolerant devices can serve as demonstration experiments or general-purpose quantum systems typically comprising tens or hundreds of qubits [10]. For example, Kim *et al*. (2023) employed a post-processing technique to mitigate errors in a noisy 127-qubit processor [11]. However, scaling to solve industrially relevant problems requires fault tolerance through QEC [12]. Moreover, QEC incurs substantial costs, significantly increasing qubit requirements by thousands and runtime by hundreds [13]. Beverland *et al*. (2022) developed a framework for quantum resource estimation and assessed three scaled quantum applications. They found that hundreds of thousands to millions of physical qubits are needed to achieve practical quantum advantage. The *quantum dynamics* simulation, *quantum chemistry* application, and *factoring* (three problems that are out of reach for classical computation) would respectively require 230, 2740, and 25481 logical qubits and up to 8.2, 6.9, and 37 millions of physical qubits [13]. Different qubit implementations might allow for a reduction of the scaling factor from physical to logical qubits. Recently, Lachance-Quirion *et al*. (2023) presented experimental results demonstrating quantum error correction of Gottesman-Kitaev-Preskill states based on reservoir engineering of a superconducting device. The lifetime of the logical qubit is increased by QEC, therefore reaching the point at which more errors are corrected than generated [14]. Other authors demonstrated a fully stabilized and error-corrected logical qubit [15]. Also, the Google Quantum AI lab has implemented a distance-7 surface code with 105 physical qubits, achieving a logical error rate below the threshold needed for exponential suppression of the logical error rate [16]. However, QEC surface codes require classical overhead for decoding syndromes [17], [18].

QEC is crucial for solving industrial-scale problems, potentially requiring more computation time, energy, and electronic components compared to initial laboratory-scale quantum experiments. While quantum computers promise advancements in sustainable information technology (IT) [19], its environmental impact is yet to be fully understood. Classical computers have been extensively studied for their environmental impact using Life Cycle Assessment (LCA), a comprehensive method evaluating systems from material procurement to end of life. This approach identifies environmental burdens and informs decision-making. For instance, a study on data centers highlighted needs for broader considerations beyond operational energy use [20]. Bol *et al*. (2011) found significant energy variations between different CPU types [21]. McDonnell (2013) analyzed a range of supercomputer designs where cooling and computing phases contribute most significantly to air acidification, while cooling manufacturing has the largest impact on ozone depletion and cancer risk [22]. Subramanian and Yung (2017) compared desktop computers and all-in-ones [23]. Maga *et al*. (2013) demonstrated lower greenhouse gases emissions in server-based computing with thin clients compared to desktop computers [24]. Loubet *et al*. (2023) conducted a comparative LCA showing significant environmental benefits of single-board computers supported by servers over traditional desktop PCs [25].

As the basis of assessment, functional units in LCAs of IT equipment typically measure operating time or product quantity [23]–[28]. Few LCA were more specific with the transmission of 1 data bit between mobile phones as the functional unit [29], [30].

A very recent master thesis conducted an LCA to estimate the environmental impacts of the production and use of a superconducting quantum computer. The total carbon footprint was estimated at around 50 t $CO_2$eq. for a Belgian energy mix and 150 t $CO_2$eq. for a global energy mix. Due to its gold components, the cryostat was identified as the subsystem with the highest impact contribution during the production phase [31]. The scope of the study includes one quantum computing chip without error correction. Although the principle of attributional LCA allows for scaling the inventory in proportion to the functional unit, considering industrial-scale calculations for over 100 logical qubits would necessitate incorporating error correction and multiplexing.

To date, no research has compared the environmental impact of quantum computers and supercomputers on an industrial scale. This study aims to create the environmental profile of a quantum computer and compare it with a cutting-edge supercomputer. Using ISO 14044:2006 guidelines, the study follows the four steps of Life Cycle Assessment (LCA): goal and scope definition, life cycle inventory (LCI), impact assessment, and result interpretation [32].

The paper is structured as follows. The Method section outlines the goal and scope of the comparative assessment, presents the LCI for both systems, and details the environmental impact assessment method and the sensitivity analyses. In the Results and Discussion section, main scenario outcomes and their sensitivity analyses are interpreted and discussed in relation to existing literature. The Conclusion summarizes key findings and implications.

# 2. Method

## 2.1. Goal and scope

The goal is to compare the potential environmental impact of a quantum vs a classical computer. We have chosen to study superconducting quantum computing architectures as they are the most prevalent. The scope of the research is to understand, according to their computing performance, how the environmental difference between both systems evolves. To do this, we must compare two machines that have similar computing capabilities. Since quantum computers and classical computers operate on fundamentally different principles, this comparison is inherently difficult. However, the purpose of quantum computers is to compute tasks that cannot be done even by the most advanced classical supercomputers of today, which we have thus chosen as a comparison point.

The scenario A of the modelling includes the four sub-systems of a cryogenic platform for superconducting quantum computing, the state-of-the-art system for running quantum computations. The data used for modeling come from the Bluefors system, first generation [33]. Four sub-systems compose the quantum computer: the cryostat, gas handling system (GHS), compressor, and control unit (CU). The scenario B is the classical supercomputer, it consists of several compute blades. The data used for modeling come largely from a Dell compute blade [34] and are scaled to the TOP500 supercomputer. A supercomputer consists of several compute blades. For our comparison, each compute blade contains two CPUs and two GPUs, as well as two power supplies and eight hard drives. Both computer lifetimes are assumed to be equal because of their 3 years warranty [35], [36]. To account for difference in the performance of the systems studied, an LCA compares them based on the functions they perform. However, the difference in the performance is unknown. Literature indicates that computing time evolves with technical and scientific breakthroughs. A functional unit with a fixed functional equivalence ratio between the systems would limit the scope of the research. Therefore, the chosen functional unit is equal operation time, and the evaluation considers several operation times.

## 2.2. Life cycle inventory

The life cycle inventory (LCI) mainly relies on the ecoinvent database [37] (ecoinvent 3.9 allocation cut-off by classification). This database is a consistent and comprehensive source of life cycle inventory data. It contains intermediate flows, i.e., processes that represent activities, systems, materials, energy sources, and wastes that belong to the Technosphere (human activities). It also contains elementary flows, i.e. resources and emissions exchanged between the Technosphere and the Ecosphere (air, water, and soils). The ecoinvent database was used or adapted to build the inventory for both computers, as they do not already exist in this database. Background data (the rest of the human activities that are connected to both systems) come from ecoinvent. For the quantum computer, foreground data, describing the system technically, mainly comes from data collection on site, technical documentation, and drawings. For the supercomputer, data comes from literature (mainly from Loubet *et al.*) and technical documentation. The life cycle inventory is broken down into four phases: the production phase (in blue in Figure 2-1) which includes resource extraction, material production, subsystem manufacturing, assembly, and intermediate transportation; the delivery phase (in orange), which covers the delivery of all subsystems; the use phase (in grey); and finally, the end-of-life phase.

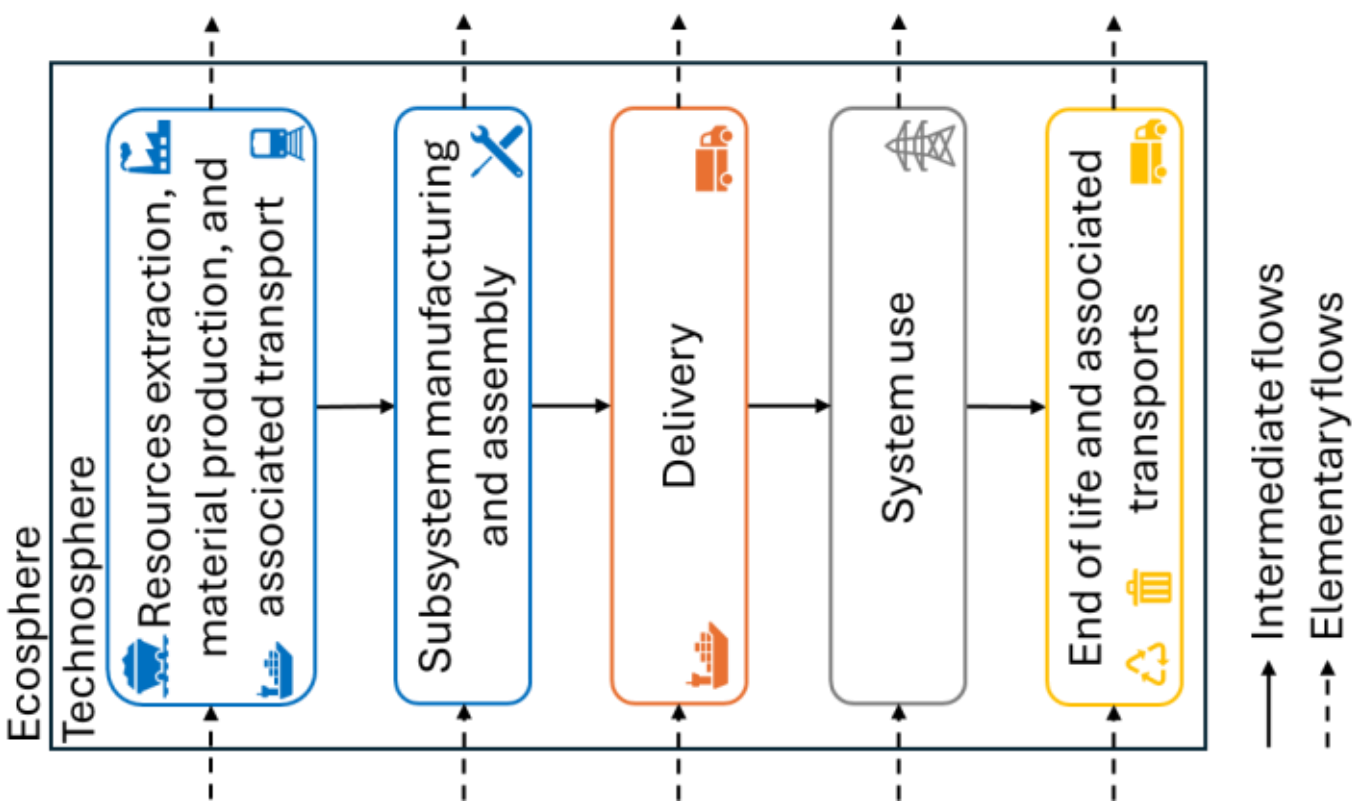

*Figure 2-1. System boundary of both systems. Production stage in blue, delivery stage in orange, use phase in grey, and end of life in yellow.*

The mass of the components was determined using the cited sources above, the estimated volume and density, or ecoinvent data. When it was not possible to determine the mass of a subpart, it was estimated based on the mass balance of the whole part. For the modeling, all the ecoinvent processes are representative of the global market, except for the use phase, which is assumed to take place in Quebec.

### 2.2.1. Production

After resource extraction and material production, manufacturing processes are considered to shape the parts. When an intermediate manufacturing process was assumed between the production of the material and machining, the intermediate process did not include the mass of the component but rather the raw mass before machining (see the supporting information for more details - Table S.I. A.2.1).

For the quantum computer, its four subsystems are modeled separately. The cryostat modeling is based on the BlueFors LD400. Regarding its gold-coated parts, the thickness assumption is 1.5 µm. The GHS is based on the GHS400, and the compressor on the CP1110. The modeled CU is of the first generation. In the room housing the cryostat, structural materials are mainly aluminum, copper, or stainless steel to avoid magnetic fields and therefore noise. Regarding the quantum error correction equipment, the assembly described by Lachance-Quirion *et al*. (2023) corresponds to one qubit. This section describes only how the inventory is built. The next section (2.2.5) explains how the configuration is scaled to 100 logical qubits, a commonly used threshold where quantum computing will be able to perform computations that classical computing cannot, especially for condensed matter physics problems [38]. The list of electronic equipment used for error correction is available in the supporting information of Lachance-Quirion *et al*. (2023). An exception is the oscillator. The modeled oscillator is the M9046A PXIe with 18 slots and two cards (M5300A PXIe RF AWG) per qubit. Since it can accommodate 16 cards, one-eighth of the oscillator is used for one qubit. Existing ecoinvent processes serve as a proxy for component modeling where possible. Otherwise, whenever available, the materials declaration or RoHS sheet (Restriction of Hazardous Substances) for the electronic component is used to model its body. If these documents are unavailable, the technical drawing, which also indicates the main material of the component body, is used instead. If neither of these sources is available, the body is modeled using a proxy, such as other similar electronic components for which information is available. The interior of the electronic components is assumed to be a printed wiring board. Supporting information details the modeling of error correction components (Table S.I. A.2.1). The supercomputer modeling is based on a server from Loubet *et al*. and Dell (2024) (see the supporting information for more details - Table S.I. B.2.1).

For both computers, the assembly is modeled similarly to that of the desktop computer from the ecoinvent database [39], but proportionally based on the mass of the equipment to be assembled. The assembly is applied to the QEC setup, to cables different from those modeled in ecoinvent, and to each subsystem (for the quantum computer: the cryostat and its support frame, the GHS, the compressor, and the CU; and also for the compute blade of the classical computer).

#### 2.2.2. Delivery

The cryostat and its support frame, the GHS, and the CU are assumed to come from Finland. The compressor is assumed to come from the USA, and its liquid nitrogen tank is sourced from France. The electronics for the QEC, as well as the associated cables—which are proportional in quantity to the QEC system—are sourced from the USA. The compute blade of the supercomputer comes from China.

#### 2.2.3. Use

For both computers, neither sleep mode nor off mode is included. However, both systems include the use of a desktop computer for data management and to start computations.

For the quantum system, primarily electricity and nitrogen refueling are required (10 liters per week per compressor). The modeling of lost nitrogen is considered an elementary flow emitted into the air. Helium required for cooling is accounted for only in the production phase because the helium circuit is closed, without loss nor refueling during the use phase. Power consumption varies depending on the number of qubits needed for the experiment, with scaling and multiplexing factors. Energy consumption of the cryostat and the electronic racks is determined based on the electronic components from Lachance-Quirion *et al*. (2023), in which some components consume active energy. To achieve one logical qubit, the electronics in the cryostat require approximately 36 W, while the rest (mostly oscillator cards in the racks) require approximately 209 W (table S.I. A.2.3 in the supporting information). For the cryostat itself, one GHS, operating with two compressors, requires 1.8 kW; one compressor requires 10.7 kW, and we assumed negligible power for the control unit according to daily uses in laboratories.

For the classical system, the supercomputer requires electricity only during the use phase. With a Dual Xeon 8168 processor of 205 W (TDP) [40] and two GPUs (e.g., an Nvidia P4000), the combined power requirement of one compute blade is 1450 W [41]. For this, two 1100 W power supply units used at 66% of their capacity are required per compute blade. In total, these supercomputers typically use thousands of those compute blades, consuming MWs of energy. It is this astronomical energy cost that prohibits scaling these systems even higher rather than hardware costs, thus supercomputers and data centers are nowadays built around energy supply constraints [7].

### 2.2.4. End of life

The end-of-life modeling uses the cut-off approach. In this system model, waste is the producer's responsibility. Additionally, recyclable products are considered burden-free because the impact of recycling is attributed to the next production [42].

### 2.2.5. Error correction and scaling the number of qubits

Lachance-Quirion *et al*. (2023) presented experimental results demonstrating quantum error correction of Gottesman-Kitaev-Preskill states based on reservoir engineering of a superconducting device. Since their experimentation applies to only one logical qubit, this section upscales it to 100 qubits. However, this is not as simple as multiplying every component by a factor of 100. Even if the experiment by Lachance-Quirion *et al*. used only 1 physical qubit to create a logical qubit, it is unlikely that this can scale to 100 logical qubits. We thus estimate that seven physical qubits are required per logical qubit (as in Steane's code [43]) and so multiply the electronics by seven. However, since electronic equipment and cables allow multiplexing, scaling also includes a division by four [44]. These two numbers, the overhead factor (O) for QEC of 7 and the multiplexing factor (M) of 4, are chosen to the best of our knowledge about technological advancements in late 2024. This scaling of O/M will undoubtedly evolve in the coming years as they are crucial to achieving realistic quantum computer sizes and energy requirements. In summary, there is a need for 175 times the setup used by Lachance-Quirion *et al*. (2023), including its associated cables, to create a quantum computer which is relevant to compare with today's supercomputers. To accommodate the extra cables and electronic components, the modeled cryostat, its support frame, and the compressor need to be multiplied by six. To operate those six cryostats, only three modeled GHS units and one CU are required. Once the system is scaled, the total power is 112.5 kW, with the compressors being the main source (64 kW), followed by the QEC setup (43 kW) in the cryostat and the racks.

This estimated energy cost for the cryostat and the racks to achieve 100 logical qubit is higher than in some other sources [7], where the main sources of power needed for a typical single QPU of 53 qubits (e.g., Google superconducting computer) are the cryostat (∼10 kW), to operate at around 15 mK, and the electronic racks (∼5 kW) for oscilloscopes, microwave electronics, analog–digital converters, and clocks. However, these setups operate physical and not logical qubits, and so do not perform QEC which accounts for much of our estimated energy consumption. Moreover, the setup studied here, based on Lachance-Quirion *et al*. (2023), performs a fully autonomous QEC of bosonic GKP states, and thus does not require any additional classical hardware.

### 2.2.6. Scaling the number of CPU cores

To compare the quantum computer to a supercomputer, the modeling of the compute blade from Loubet *et al*. (scenario B) serves as a basis for comparison and is scaled to a TOP500 supercomputer [45], [46] with a 64-core 2 GHz processor, using the number of processing cores as scaling parameter. Table 2-1 displays the key scaling parameters relative to the number of CPU cores. 12,630 compute blades are compared to the quantum computer.

*Table 2-1. Summary of the orders of magnitude of the modeled compute blade compared to a reference supercomputer*

| | **Dell Precision 7920 compute blade (scenario B)** | | **Frontier - HPE Cray EX235a (TOP500)** | |
|---|---|---|---|---|
| Total power usage (kW) | 18,312.53 | * | 21,000 | [46] |
| Power/Compute blade (kW) | 1.45 | [41] | 4.43 | * |
| CPU/compute node | 1 | | 1 | [47], [48] |
| Compute node/compute blade | 2 | [25], [41] | 2 | [46], [47] |
| Chassis/cabinet | | | 8 | [49] |
| Compute blade/chassis | | | 8 | [49] |
| Compute blade/cabinet | | | 64 | [49] |
| Cores/CPU | 24 | [25], [36], [40] | 64 | [48] |
| CPU/cabinet | | | 128 | [49] |
| Cabinet | | | 74 | [47] |
| Total CPU | 25,258.67 | * | 9,472 | * |
| CPU/compute blade | 2 | [25], [41] | 2 | [47] |
| **Total Cores CPU (processor)** | **606,208** | * | **606,208** | * |
| **Total Compute blades** | **12,629.33** | * | 4,736 | * |
| Masse/Compute blade (kg) | 28.6 | [34] | 57 | * |
| Total masse (kg) | 361,199 | * | 268,546 | * |

* Figures obtained by calculations

With this scaling, the modeled server is expected to have 167% more compute nodes than the TOP500 supercomputer and be 35% heavier. However, its estimated power consumption is 13% lower.

As mentioned in the introduction, the challenge related to the difference in computation time between quantum and classical computers is that advances in classical methods can also reduce this gap. Assuming the same lifespan for both types of computers, the comparison will be made with respect to the same processing time at multiple instances. This comparison will help understand how the difference in potential impacts evolves between the two state-of-the-art computers depending on the calculation time. The comparative scenarios are as follows: A) Quantum computer, with QEC to reach 100 logical Qubits, B) Classical computer, with 606,208 CPU cores.

### 2.3. Environmental impact modeling

For this study, the IMPACT World+ [50] methodology (Expert 2.0.1 version) was selected. It characterizes the inventory of the elementary flows into 21 specific damage indicators. It subsequently makes it possible to group these indicators into 2 areas of protection: human health and ecosystems. To present the Climate change damage indicator separately, this subcategory is excluded from the subcategories included in the two ImpactWorld+ protection domains, and the IPCC 2021 impact calculation methodology is selected [51]. The Climate change indicator is expressed as tonne $CO_2$ eq., a unit created to compare the impacts of the different greenhouse gases (GHGs) on global warming and to be able to cumulate their emissions. The GHGs are distinguished by the amount of Infrared radiation they can absorb and their lifetime in the atmosphere. The Ecosystems indicator is expressed in Potentially Disappeared Fraction (PDF) of species over a certain amount of $m^2$ during a certain amount of year (PDF.$m^2$.y) which relates to the likelihood of species loss. The Human health indicator is expresses as Disability-Adjusted Life Years (DALY). DALYs represent years of life lost due to premature death plus years of healthy life lost due to disability or disease for a given number of individuals. The software that allows to manage the inventory and the characterization of impacts is SimaPro 9 [52].

### 2.4. Sensitivity analysis

In quantum computing, the efficiency of error correction may vary from one experiment to another due to noise, and multiplexing may depend on the hardware. These two parameters, resulting from assumptions, are therefore subject to sensitivity analysis (Table 2-2). The sensitivity analysis scenario A' uses the Lachance-Quirion *et al*. setup 2500 times instead of 175 times, which is closer to typical assumptions of supercomputing qubits that do not use bosonic GKP codes but rather a surface code [53]. These architectures do not use the same scaling rules as bosonic codes, and the multiplexing and cryostat scalings have been adjusted accordingly. However, these exact numbers are hypotheses that will need to be adapted in the future, as quantum hardware providers discover the exact constraints of scaling their particular architectures. Indeed, the ratio of 1000:1 in scenario A' is taken as an upper bound encompassing multiple layers of encoding. A scenario with a 100:1 ratio would assume that a single layer of encoding is needed for fault-tolerant operation.

*Table 2-2. Quantity of subsystems for 100 logical Qubits (# / cryostat)*

| **Scenario** | | **A** | **A'** |
|---|---|---|---|
| Parameters | Logical qubits | 100 | 100 |
| | Physical qubits per logical qubits | 7 | 1000 |
| | Multiplexing | 4 | 20 |
| Subsystems | Cryostat and its support frame | 6 | 10 |
| | - QEC setup and its assumed cables | 175 (29.17*) | 5000 (500*) |
| | Gas Handling System | 3 | 5 |
| | Cryocooler System | 6 | 10 |
| | Control Unit | 1 | 2 |
| * Quantity per cryostat | | | |

In scenario A, classical hardware needed for decoding syndromes during QEC is considered [14]. In scenario A', where the conservative 1000:1 ratio would be representative of a QEC surface code, classical overhead for decoding syndromes is not taken into account. No data was available to base a thorough modelling of such QEC classical hardware, in part because current experiments have very recently been capable of creating logical qubits and only use QEC on logical qubits used as memories and not on logical qubits used for computation.

## 3. Results and discussion

The results of scenarios A (quantum computer with autonomous QEC of bosonic codes) and B (classical computer) are presented in section 3.1, and the sensitivity analysis in 3.2.

## 3.1. Comparison with a supercomputer

Figure 3-1 displays the relative contributions of the life cycle phases to Climate change (t $CO_2$eq.) for each computer. It shows that the production phase is the major contributor for both computers up to 10,000 hours of use (a little over one year). Beyond this point, the use phase becomes the most significant contributor. At 50,000 hours of use (close to 6 years), the use phase remains the second most significant phase in scenario A, while it becomes the most contributing phase in scenario B. Finally, the delivery phase is equivalent to the end-of-life phase in scenario A but is more important in scenario B.

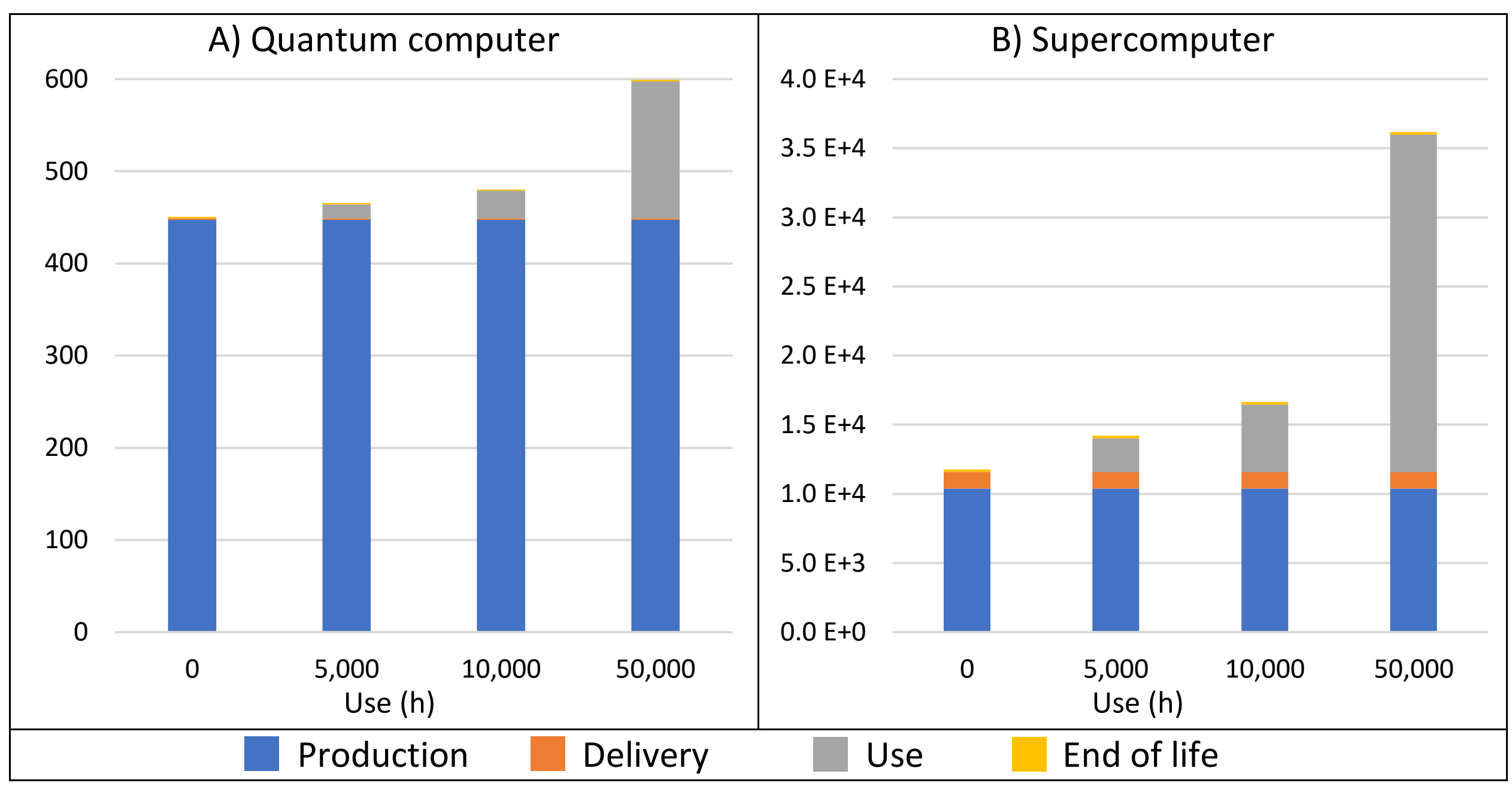

*Figure 3-1. Contribution of the life cycle phases to climate change impact (t $CO_2$eq.).*

The difference in impact between the two computers is two orders of magnitude. It is mainly explained by the power and materials used: 18,312 kW and 380,768 kg (Table 2-1 + cables) for the supercomputer versus 112.5 kW and 9,622 kg for the quantum computer (see the supporting information : Table S.I. B.2.3, Table S.I. B.2.1, Table S.I. A.2.3, and Table S.I. A.2.1).

Loubet *et al*. estimated the usage time (40% at 201.4 W), sleep time (5% at 11.4 W), and off time (55% at 0.3 W) for their server. Their estimated annual energy consumption is therefore 712.1 kWh/year, which represents 20% of the climate change impact over a lifetime of 5 years with a French electricity mix [25]. The annual energy consumption of the supercomputer modeled for scenario B is calculated as 1.45 kW * 8760 hours/year = 12,702 kWh/year. As presented in Figure 3-1, the use phase would account for around 65% at a usage time of 5 years with a Quebec electricity mix. The difference in energy consumption can explain the difference in the contribution of the use phase. Also, the difference between the data sources can explain the discrepancy between the annual energy consumption of Loubet *et al*. and scenario B. They used the 7920 tower specification [54], [55], but no details were found on the relationship between power requirements and the number of CPUs and GPUs. Scenario B considers the combined power requirements of two CPUs and two GPUs [41] (giving a total power usage close to the TOP500 supercomputer – see Table 2-1). Moreover, not considering sleep time or off time in the modeling can explain the difference compared to the fragmented use in the literature. However, since both scenarios A and B include a continuous use phase, the results remain consistent. The annual energy per compute blade of scenario B may seem astronomical compared to Loubet *et al*., this is because the usage of supercomputers in scientific high-performance conditions is almost perpetual.

Regarding the quantum computer, over a 5-year lifetime, the modeling by Billat and Doeran (2024) estimates a climate change impact of 50 t $CO_2$eq. for a Belgian energy mix and 150 t $CO_2$eq. for a global energy mix. This is lower than what is shown in Figure 3-1. A) (approximately 583 t $CO_2$eq. at 43,800 hours) and is to be expected due to the difference in scale with this paper's modeling of logical qubits accounting for error correction. Table S.I. A.3.2 in the supporting information shows the relative contributions of the subsystems to climate change. The production of one cryostat contributes more to climate change than that of the GHS and the compressor, which have equivalent contributions. The CU production contributes the least. This trend is also reflected in the modeling by Billat and Doeran (2024) [31]. However, the absolute impacts of the subsystems are higher than those reported by Billat and Doeran (2024). Since gold is a significant contributor, this difference probably is due to the modeled gold coating of 1.5 µm compared to 1 µm thick for Billat and Doeran (2024). Additionally, the cryostat seems to contain more electronic components (due to the QEC), and the modeling seems to include more manufacturing processes, in contrast to the coarse-grained modeling of Billat and Doeran (2024). Regarding the use phase, it represents less than 25% of impact (Figure 3-1. A) at about 43,800 hours) which is less than in the Billat and Doeran modeling (80% of the impact). The identifiable reasons may be due to the more impactful production modeling per subsystem in our study, also including more quantum hardware per cryostat because of the QEC. Another identifiable reason would concern the power of the GHS and CU subsystems, lower than those of Billat and Doeran (see Table S.I. A.2.3').

Figure 3-2 shows the relative contributions of the life cycle phases to the three impacts for each computer on different axes. The same trend of the climate change results is observed with a difference of two orders of magnitude for the Ecosystem indicator and one for Human health. A particular aspect of the Ecosystems indicator is that the use phase becomes more important than the production phase earlier compared to the other two indicators. During this phase, electricity consumption is the single most dominant contributor. In a region where electricity consumption has a greater impact, and in the hypothesis of non-stop industrial use, the dominance of its contribution would occur earlier, closer to the three years guaranteed for the quantum computer.

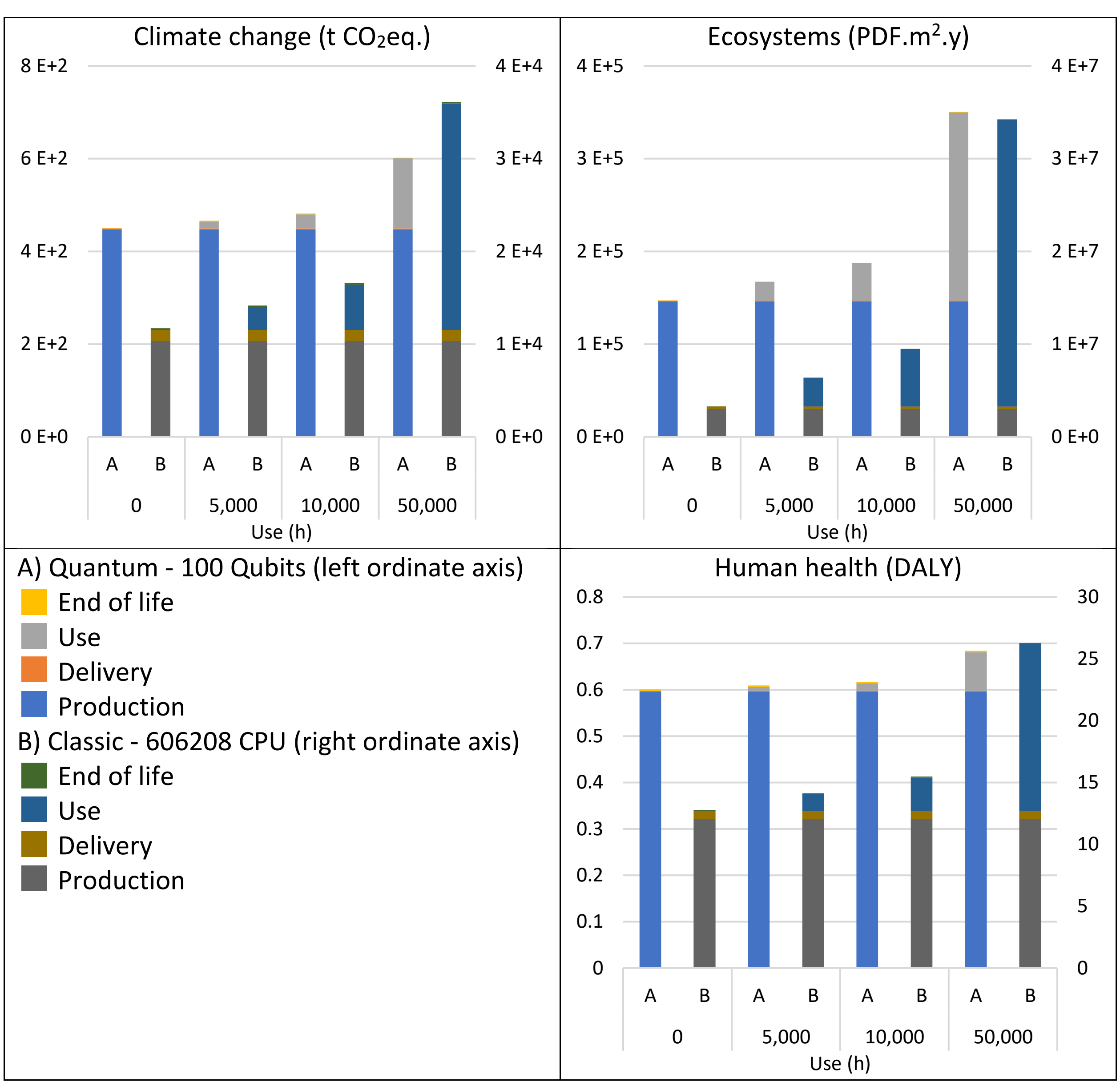

*Figure 3-2. Comparison between the quantum and the supercomputer according to duration of use.*

In the production phase of the quantum computer, the cryostat contributes the most to all three indicators, accounting for between 89% and 92%, depending on the indicator (Figure A-1. A). The primary reason is the experimental part involving QEC, which contributes between 59% and 64% of the cryostat's impact, according to the indicators (Figure A-2. A). The estimated weight of the QEC components is 10 kg per qubit. Therefore, scaling parameters play a crucial role in modeling the production impacts. For scenario A, it is estimated that 175 times the setup described by Lachance-Quirion *et al*. is needed, equivalent to 1750 kg of material (excluding cables). The second most impactful aspect of cryostat production is the insulating envelopes. Although these are made of aluminum, the inner envelope of the experimental space is entirely covered in gold. This gold envelope is optional, but it is included in the modeling for conservatism. In gold production, the refinery is the largest contributor. The third most impactful part is the structure of the cryostat chambers, particularly those with the highest amount of gold coating.

In the production phase of the classic computer, the motherboards with CPUs (approximately 30%), processors (about 21%), and power supplies (about 12%) contribute the most (Figure A-3). The two GPUs account for around 7%.

The delivery phase is clearly more impactful for the supercomputer than for the quantum computer, as the total mass-kilometer is greater for the supercomputer (6,574,899 t.km vs. 41,310 t.km). Although the end-of-life impact is not very apparent due to the cut-off approach applied to electronic components, the difference in mass at the end of life explains the difference in impact for this phase.

## 3.2. Sensitivity analysis

Figure 3-3 shows the sensitivity analysis regarding the number of physical qubits and the multiplexing based on the duration of use. Regardless of the environmental indicator, the quantum computer remains less impactful than the modeled supercomputer. The impacts of the production of the quantum system with a ratio representative of a surface code (A') are around 86% and 95% of those of the classical system. The cryostat and QEC system are the only major contributors (99 % - Figure A-1. A’) to the production phase. This is due to the QEC equipment, which is increased by a factor of 28.57 from scenario A to A’, while the cryostat, the GHS, and the compressor are multiplied by 1.6, and the CU by a factor of 2. Regarding the usage phase, the fact that the QEC equipment becomes more energy-intensive than the compressor in this scenario (1,227 kW versus 107 kW) explains why it appears more prominent earlier. Furthermore, QEC of a surface code requires classical overhead for decoding syndromes. This overhead was not taken into account because of data availability and since current experiments only use QEC on logical memories and not on logical qubits used for computation. However, recent work in this field showed that decoding hardware would consume power in the order of mWs to Ws using dedicated Field Programmable Gate Arrays or Application Specific Integrated Circuits [56], [57], which would not have a major impact on our analysis. Further analysis of the impact of classical QEC hardware at scale is an important avenue of research to determine the validity of those estimates in future fault-tolerant quantum computers. These considerations will also greatly vary depending on architecture, so further LCAs are needed for photonic, trapped ions and neutral atoms quantum computers.

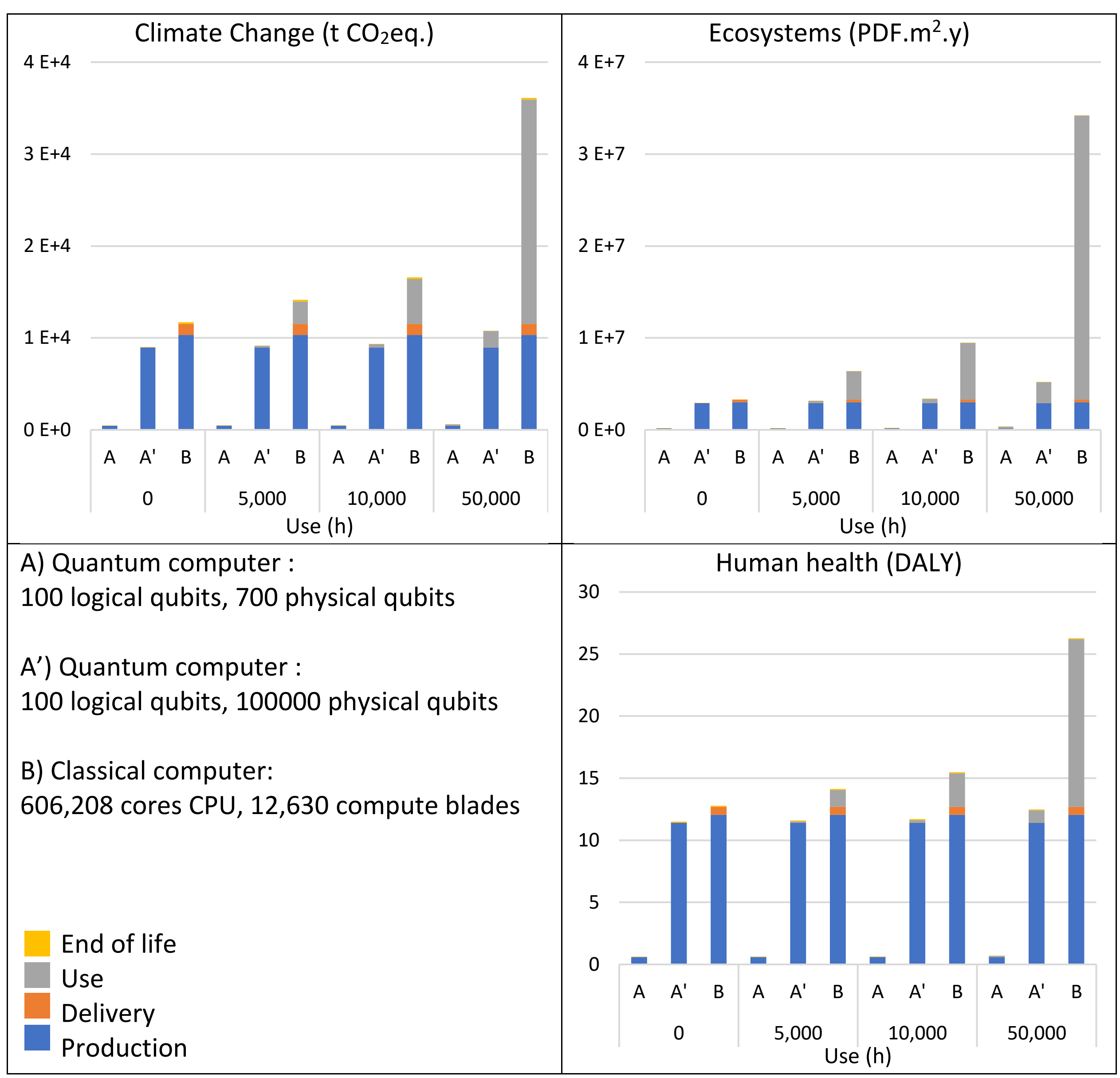

*Figure 3-3. Sensitivity analysis on the number of qubits.*

The impact of production, transport, and end-of-life for scenario A' is 77%, 88%, and 90% of that for scenario B in the categories of Climate Change, Ecosystems, and Human Health. Additionally, considering the use phase, these percentages will decrease as the time of use increases. Therefore, given the same lifetime and computation time, the configuration of the superconducting quantum computer with 100 logical qubits has a better environmental potential than a classical supercomputer.

Since the supercomputer model used as a baseline is less optimal than a TOP500 supercomputer (24 cores per CPU and 2 CPUs per compute node versus 64 cores per CPU and 2 CPUs per compute node, see Table 2-1), it can be assumed that the production, delivery, and end-of-life impacts of scenario B are overestimated compared to the TOP500 supercomputer. Indeed, the modeling includes 12,630 compute blades versus 4,736 for the TOP500 supercomputer. If the modeled computer, used for comparison, had 64-core CPUs, then 4,736 computing blades would be modeled, but the motherboard modeling would need to be scaled up accordingly. Even with this reduction of approximately a factor of 2.7 (which does not take into account possible differences in production or energy use of the CPUs with more cores), its impact remains at least an order of magnitude over the quantum computer for scenario A at all times (see scenario B' in Figure A-4). However, compared to scenario A', the impact of the classical computer is higher after 50,000, 10,000 and 85,000 hours for the Climate change, Ecosystems, and Human health indicator. This shows that it is crucial to consider the scaling factors of quantum computers in order for them to be more energy efficient. Finally, it should be mentioned that even if a part of the cooling systems is modeled with fans in the compute blade (Table S.I. B.2.1), the supercomputer room's cooling systems is missing. If added to the modeling, it would increase the estimated impact of the classical computer. This should increase the impact differences between A and B. Regarding scenario A', it is not possible to conclude because its environmental advantage depends on the hours of use and perhaps the resource needed for decoding.

## 3.3. Limits

Before concluding, it is important to list the limitations of the model. Regarding the data, the model is composed of ecoinvent generic processes, except for the electrical mix in the use phase. Additionally, the data on electronic components from ecoinvent were collected in the 2000s [58]. It is also challenging to differentiate electronic components of the same type but with different performances, as detailed in the Green-Algorithms calculator [59], [60] which presents energy consumption based on the number of cores of the CPU and/or GPU and the available memory. This would help to differentiate environmental impact of the supercomputers of the TOP500. Furthermore, since the occupied area is not considered in the study, the associated impacts as well as the size of the buildings required to house the systems are not presented.

As presented in the introduction, the performance of both quantum and classical computers continues to improve [6]. Therefore, a comparison based on a specific computing time may be subject to change with technological advancements. Additionally, a more specific functional unit for comparison could be the PFlop [7] or, even better, specific useful algorithms run on both types of computers that achieve the same result, but those are yet to be achieved on the quantum side. To go further, more analysis could be performed on the quantity of logical qubits.

Finally, the lifespan of computers and their components is not addressed. Thus, the impacts could be greater if components need to be replaced. As technology evolves, components become obsolete in favor of improved performance. Assuming that the overall lifetimes of the two systems are equal (3 years = 26,280 hours), the impacts of production, transport, and end of life would have to be added every 3 years (however, real life experience shows that electronic components typically last for a few decades). If, in addition, the calculation time is the same, the conclusions remain unchanged. The difference in impacts between the two systems is simply amplified by the differences in the impacts of production, transport, and end-of-life. On the other hand, if the calculation time is different, a substantial difference in calculation time would be required to alter the conclusions. This difference must be such that the impact of using the quantum computer compensates for the difference in impact between the quantum production and the classical production and use. The extent of this difference is difficult to determine from Figure 3-3 because the histograms for production, transport, and end-of-life are not presented according to the duration of use and the pro rata of the lifespan.

To achieve «sustainable quantum computing» [61] and to improve the knowledge provided by our comparative LCA, further and continuous studies, such as social LCAs, life cycle assessment of costs, and other environmental LCAs, are needed. As an example, our current assessment is a relevant starting point for future more thorough LCAs taking into account the QEC decoding and correcting overhead, which will be feasible once quantum hardware reaches a state where QEC on multiple qubits is achieved and good modelling data becomes available.

# 4. Conclusion

This article compares the environmental impact of superconducting quantum computers and classical supercomputers on an industrial scale and shows how quantum computers could be less impacting than classical computers. To do this, it draws up an environmental LCA of a superconducting quantum computer including either autonomous QEC of a bosonic code or QEC of a surface code. The main results highlight the significant direct and indirect contributions of QEC equipment to the overall impacts of the quantum computer, due to the large number of electronic components required to achieve 100 logical qubits, an important threshold where quantum computers will be able to perform computations classical computers cannot. Regarding the comparison between the two computers, assuming equal lifetimes, the number of computing blades and their total electrical consumption make the classical supercomputer more impactful in the three environmental categories considered (Climate change, Ecosystems, and Human health). This is however not systematic before a certain period of use of a quantum computer using a surface code for QEC. Further LCAs and quantum experiments would help to better clarify the environmental difference. Although the two main limitations are related to the lifetime parameter and precise functional equivalence, our study demonstrates how usage time drives an environmental advantage for quantum computers over classical ones and shows that the impact of quantum computing greatly depends on the chosen architecture and QEC code. Beyond the friendly competition over the technical performance of quantum vs classical computers, this article explores underlying aspects of sustainable quantum computation and broadens the discussion on the advantages and disadvantages of quantum computers.

# Acknowledgements

We thank Christian Lupien, Max Hofheinz, Nicolas Bourlet, Gabriel Laliberté and Édouard Pinsolle from the Institut quantique for their help in gathering on-site data on the various systems necessary to run a quantum computer. We thank Dany Lachance-Quirion and Michel Pioro-Ladrière of Nord Quantique for their assistance with the QEC setup, electronic components, and scaling. We also thank Alain Veilleux from the Scientific computation center of UdeS for insights regarding the use of supercomputers. This research was undertaken thanks to funding from the Canada First Research Excellence Fund.

# A. Appendix.

*Table A-1. Orders of magnitude of the modeled compute blade compared to a reference supercomputer*

| | Dell Precision 7920 compute blade (scenario B) | | Frontier - HPE Cray EX235a (TOP500) | |
|---|---|---|---|---|
| Total power usage (kW) | 18,312.53 | * | 21,000 | [46] |
| Power/Compute blade (kW) | 1.45 | [41] | 4.43 | * |
| CPU/compute node | 1 | | 1 | [47], [48] |
| GPU/compute node | 1 | | 4 | [47], [62] |
| Compute node/compute blade | 2 | [25], [41] | 2 | [46], [47] |
| Chassis/cabinet | | | 8 | [49] |
| Compute blade/chassis | | | 8 | [49] |
| Compute blade/cabinet | | | 64 | [49] |
| Compute node/cabinet | | | 128 | * |
| Cores/CPU | 24 | [25], [36], [40] | 64 | [48] |
| Cores/GPU | 1,792 | [63] | 213.62 | * |
| CPU/cabinet | | | 128 | [49] |
| GPU/cabinet | | | 512 | [49] |
| Cabinet | | | 74 | [47] |
| Combined CPU and GPU cores | 45,869,738.67 | * | 8,699,904 | [45] |
| Total CPU | **25,258.67** | * | 9,472 | * |
| Total GPU | 25,258.67 | * | 37,888 | * |
| CPU/compute blade | 2 | [25], [41] | 2 | [47] |
| GPU/compute blade | 2 | [25], [34], [41] | 8 | [47] |
| Compute node | 25,258.67 | * | 9,472 | * |
| Total Cores CPU (processor) | **606,208** | * | 606,208 | * |
| Total Cores GPU | 45,263,530.67 | * | 8,093,696 | * |
| Total Compute blades | **12,629.33** | * | 4,736 | * |
| RAM/CPU (GB) | 768 | [40] | 512 | [47] |
| RAM/GPU (GB) | 8 | [63] | 128 | [47] |
| Total RAM CPU | 19,398,656 | * | 4,849,664 | * |
| Total RAM GPU | 202,069.33 | * | 4,849,664 | * |
| Masse/Compute blade (kg) | 29 | [34] | 57 | * |
| Masse/cabinet (kg) | | | 3,629 | [64] |
| Total masse (kg) | 361,199 | * | 268,546 | * |

* Figures obtained by calculations

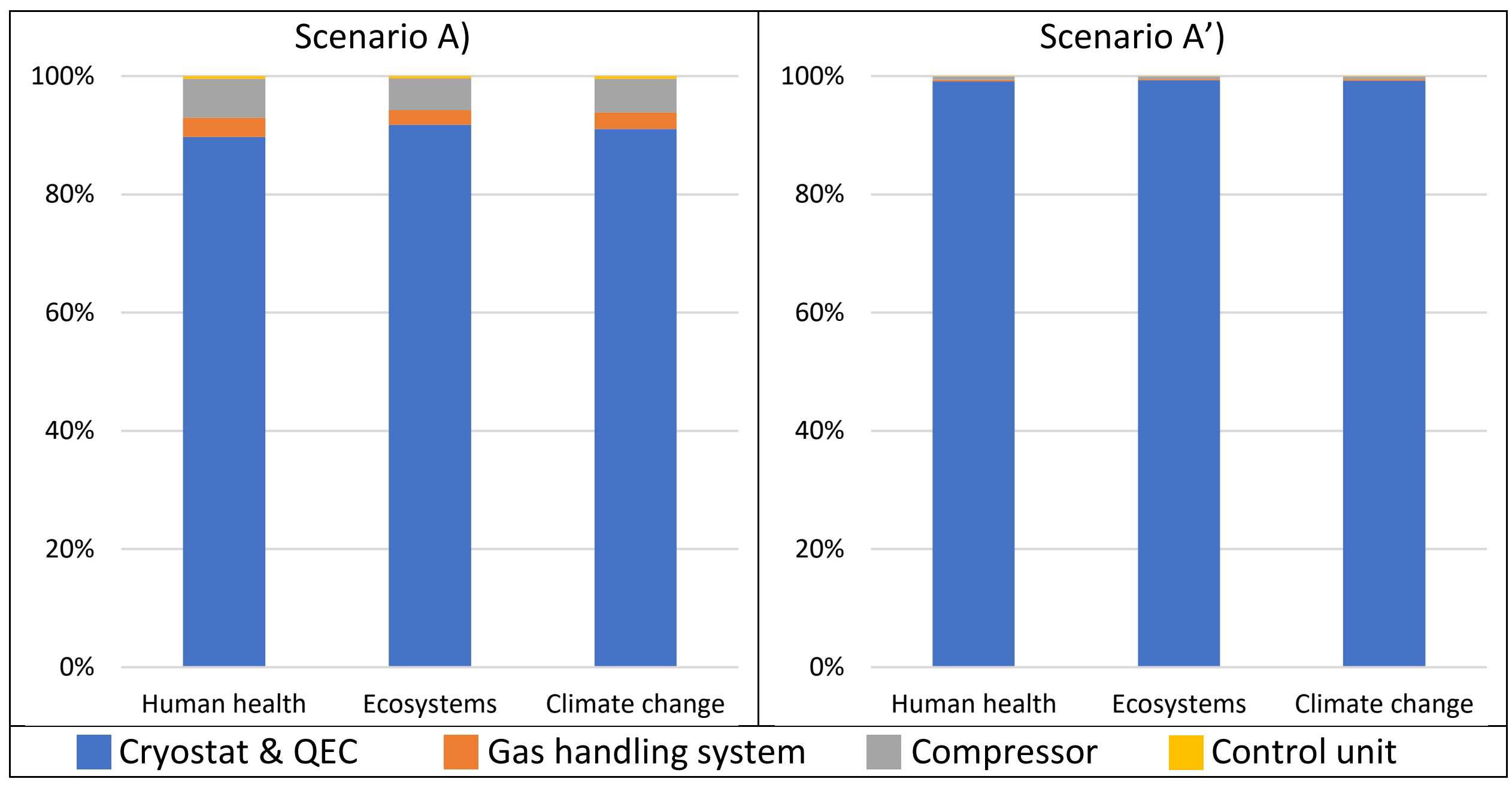

*Figure A-1. Contribution of the four subsystems of the quantum computer to the impacts of the production phase; A) Correction of 700 qubits and multiplexing of 4, A') Correction of 1,000,000 qubits and multiplexing of 20.*

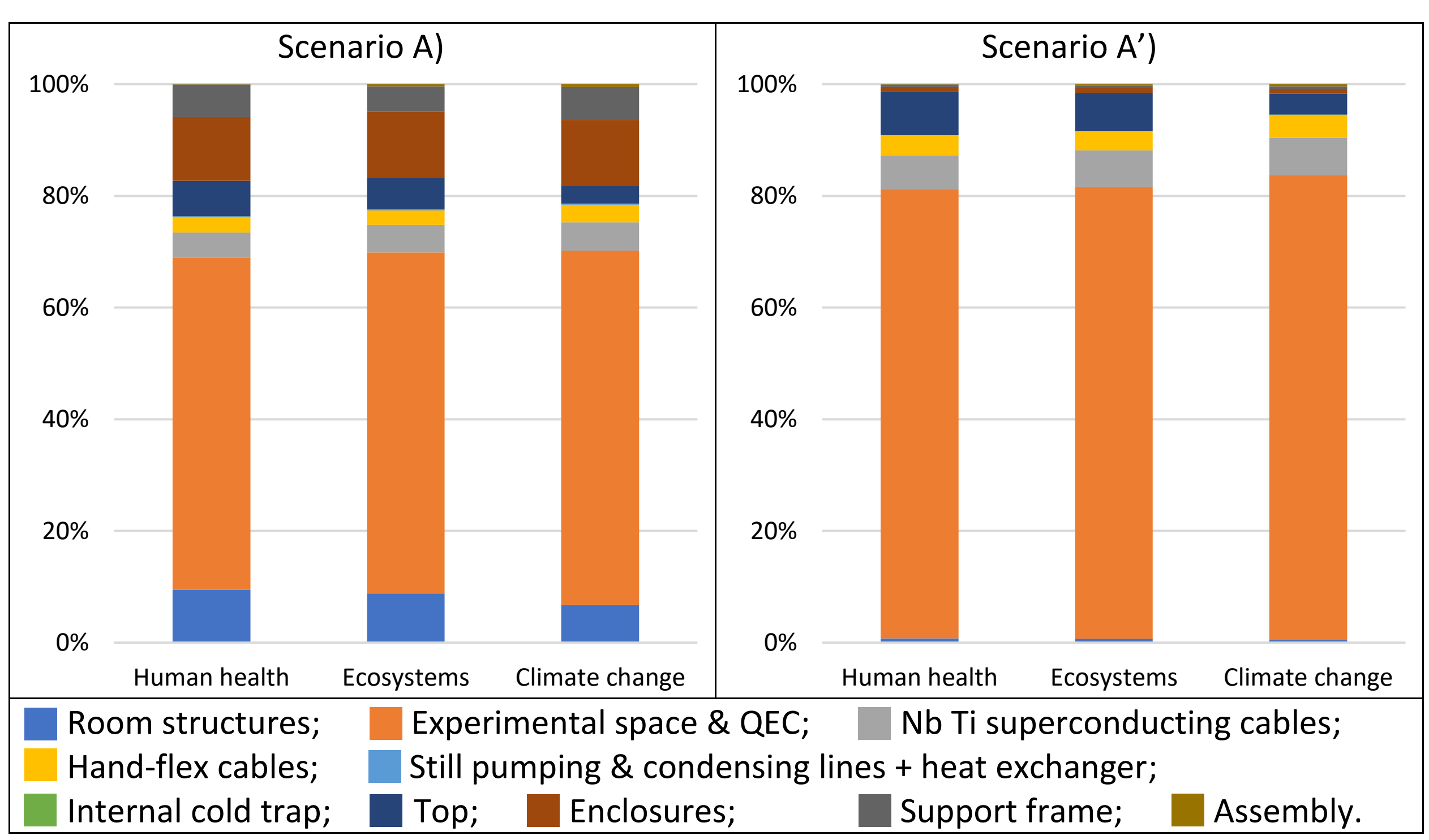

*Figure A-2. Contribution of the parts of the cryostat to the impacts of its production; A) Correction of 700 qubits and multiplexing of 4, A') Correction of 1,000,000 qubits and multiplexing of 20.*

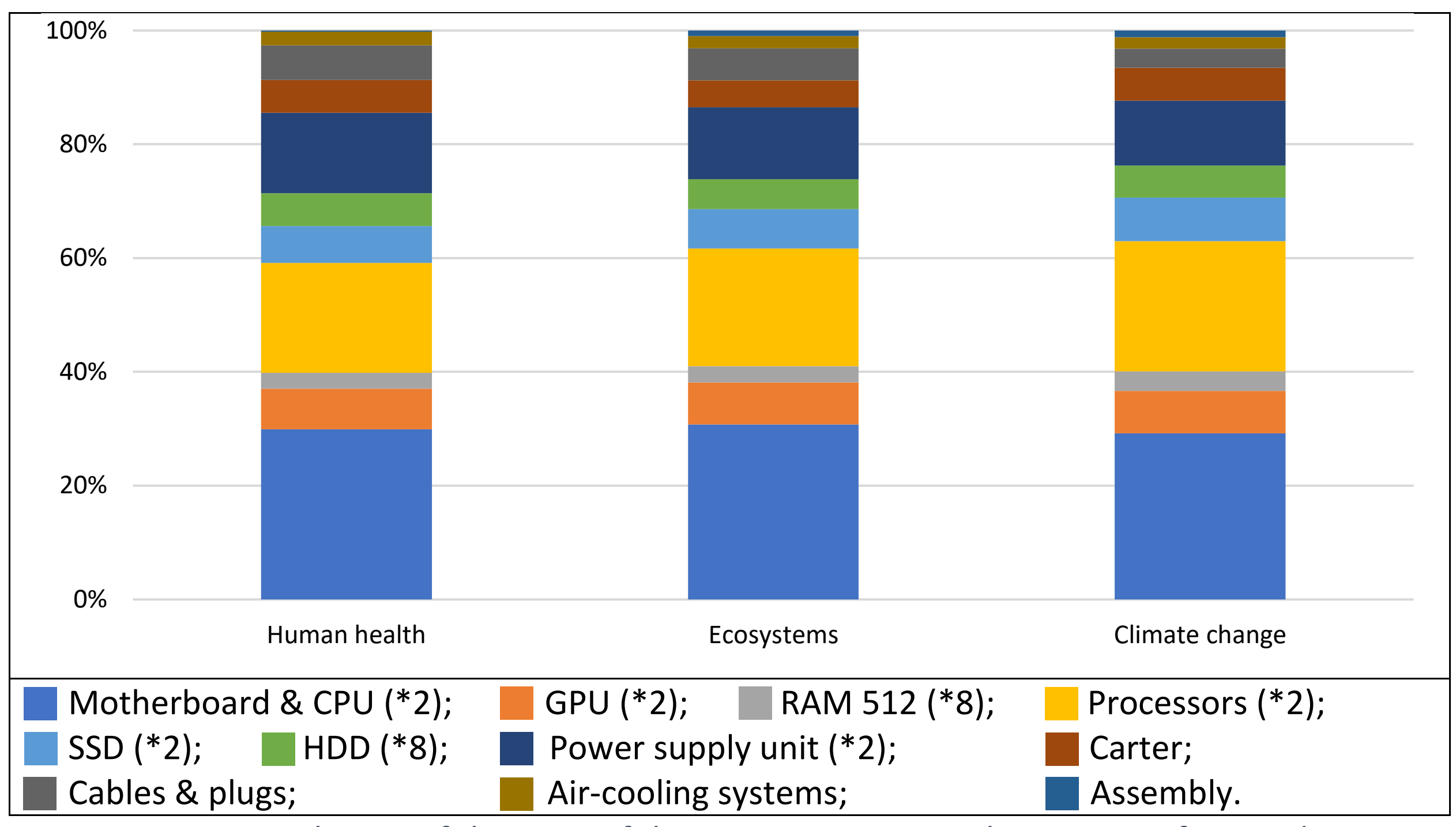

*Figure A-3. Contribution of the parts of the Supercomputer to the impacts of its production.*

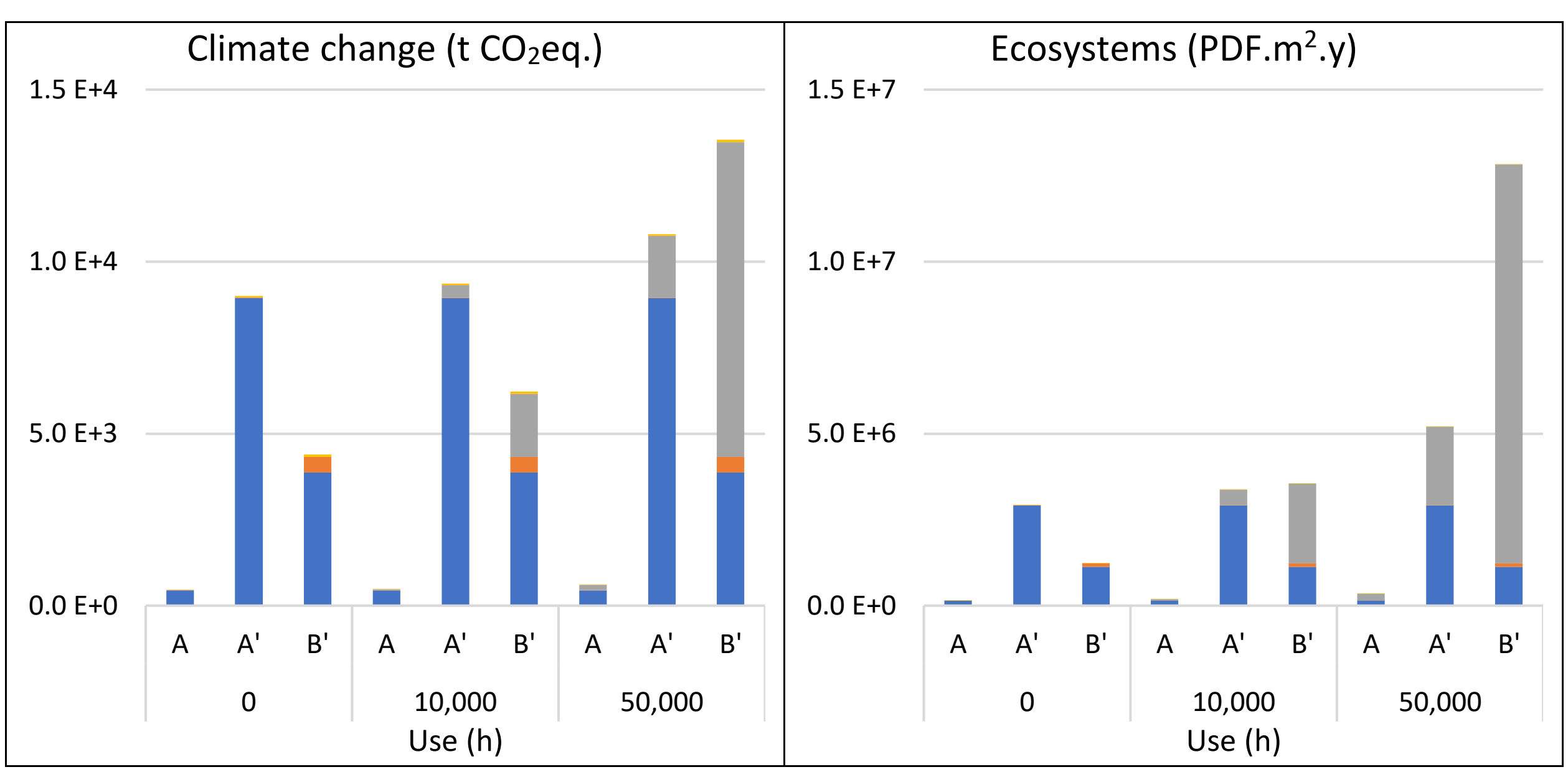

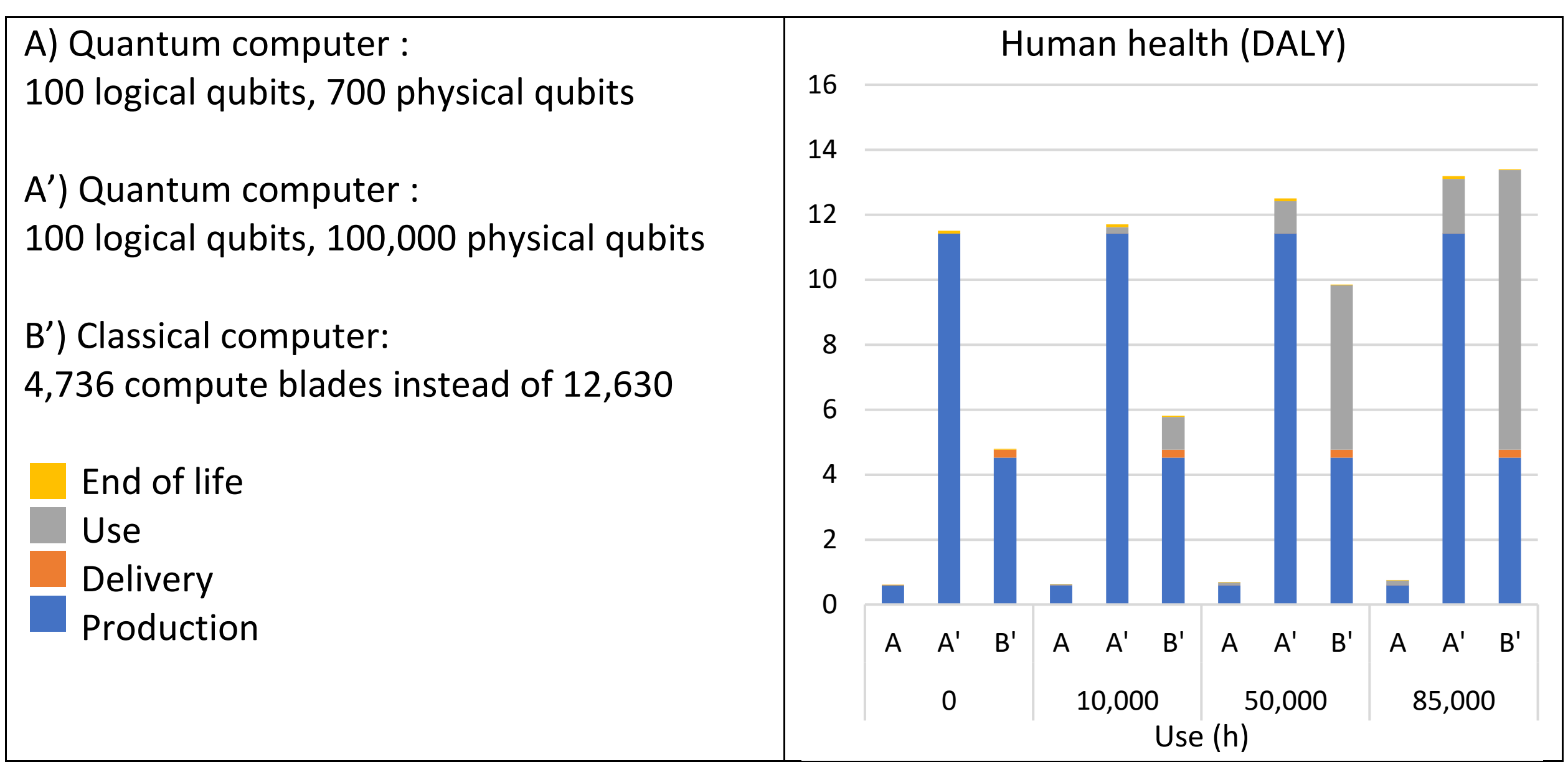

*Figure A-4. Sensitivity analysis on the number of qubits and the number of compute blades.*

## References

[1] F. Arute *et al.*, "Quantum supremacy using a programmable superconducting processor," *Nature*, vol. 574, no. 7779, pp. 505–510, Oct. 2019.

[2] E. Pednault, J. A. Gunnels, G. Nannicini, L. Horesh, and R. Wisnieff, "Leveraging Secondary Storage to Simulate Deep 54-qubit Sycamore Circuits," Preprint, last revised 22 Oct 2019 (version v2), Oct. 2019.

[3] C. Huang *et al.*, "Classical Simulation of Quantum Supremacy Circuits," Preprint, submitted on 14 May 2020, May 2020.

[4] F. Pan, K. Chen, and P. Zhang, "Solving the sampling problem of the Sycamore quantum circuits," *Physical Review Letters*, vol. 129, no. 9, Nov. 2021.

[5] L. S. Madsen *et al.*, "Quantum computational advantage with a programmable photonic processor," *Nature*, vol. 606, no. 7912, pp. 75–81, Jun. 2022.

[6] T. L. Scholten *et al.*, "Assessing the Benefits and Risks of Quantum Computers," Preprint, last revised 13 Feb 2024 (v2), Jan. 2024.

[7] B. Villalonga *et al.*, "Establishing the Quantum Supremacy Frontier with a 281 Pflop/s Simulation," *Quantum Science and Technology*, vol. 5, no. 3, p. 14, May 2020.

[8] K. A. Britt, F. A. Mohiyaddin, and T. S. Humble, "Quantum accelerators for high-performance computing systems," in *2017 IEEE International Conference on Rebooting Computing, ICRC 2017 - Proceedings*, 2017, vol. 2017-Janua, pp. 1–7.

[9] K. Bharti *et al.*, "Noisy intermediate-scale quantum (NISQ) algorithms," *Reviews of*

*Modern Physics*, vol. 94, no. 1, pp. 1–91, Jan. 2021.

[10] A. D. Córcoles *et al.*, “Challenges and Opportunities of Near-Term Quantum Computing Systems,” *Proceedings of the IEEE*, vol. 108, no. 8, pp. 1338–1352, Oct. 2019.

[11] Y. Kim *et al.*, “Evidence for the utility of quantum computing before fault tolerance,” *Nature*, vol. 618, no. 7965, pp. 500–505, Jun. 2023.

[12] G. Wendin and J. Bylander, “Quantum computer scales up by mitigating errors,” *Nature*, vol. 618, no. 7965, pp. 462–463, Jun. 2023.

[13] M. E. Beverland *et al.*, “Assessing requirements to scale to practical quantum advantage,” Preprint, submitted on 14 Nov 2022, Nov. 2022.

[14] D. Lachance-Quirion *et al.*, “Autonomous quantum error correction of Gottesman-Kitaev-Preskill states,” Preprint, last revised 19 Oct 2023 (this version, v2), Oct. 2023.

[15] V. V. Sivak *et al.*, “Real-time quantum error correction beyond break-even,” *Nature*, vol. 616, no. 7955, pp. 50–55, Apr. 2023.

[16] R. Acharya *et al.*, “Quantum error correction below the surface code threshold,” *Nature*, p. 27, Aug. 2024.

[17] J. Bausch *et al.*, “Learning high-accuracy error decoding for quantum processors,” *Nature*, vol. 635, no. 8040, pp. 834–840, 2024.

[18] C. Jones, “Improved accuracy for decoding surface codes with matching synthesis,” Preprint, Submitted on 22 Aug 2024, Aug. 2024.

[19] M. Gloukhovtsev, “How quantum computing contributes to achieving sustainable IT,” in *Making IT Sustainable*, Technique., M. Conner, H. Dwivedi, F. Fathima, and G. Harris, Eds. Elsevier, Mica Haley, 2024, pp. 251–272.

[20] A. Shah, C. Bash, R. Sharma, T. Christian, B. J. Watson, and C. Patel, “Evaluating life-cycle environmental impact of data centers,” *Journal of Electronic Packaging, Transactions of the ASME*, vol. 133, no. 3, pp. 1–9, Sep. 2011.

[21] D. Bol, S. Boyd, and D. Dornfeld, “Application-aware LCA of semiconductors: Life-cycle energy of microprocessors from high-performance 32nm CPU to ultra-low-power 130nm MCU,” in *Proceedings of the 2011 IEEE International Symposium on Sustainable Systems and Technology, ISSST 2011*, 2011, pp. 1–6.

[22] M. McDonnell, “Supercomputer Design: An Initial Effort to Capture the Environmental, Economic, and Societal Impacts,” *Chemical and Biomolecular Engineering Publications and Other Works*, p. 20, May 2013.

[23] K. Subramanian and W. K. C. Yung, “Life cycle assessment study of an integrated desktop device -comparison of two information and communication technologies: Desktop computers versus all-in-ones,” *Journal of Cleaner Production*, vol. 156, pp. 828–837, Jul. 2017.

[24] D. Maga, M. Hiebel, and C. Knermann, “Comparison of two ICT solutions: Desktop PC versus thin client computing,” *International Journal of Life Cycle Assessment*, vol. 18, no. 4, pp. 861–871, May 2013.

[25] P. Loubet, A. Vincent, A. Collin, C. Dejous, A. Ghiotto, and C. Jego, “Life cycle assessment of ICT in higher education: a comparison between desktop and single-board computers,” *International Journal of Life Cycle Assessment*, vol. 28, no. 3, pp. 255–273, Mar. 2023.

[26] F. Görkem Üçtuğ and T. Can Ünver, “Life cycle assessment-based environmental impact analysis of a tier 4 data center: A case study in Turkey,” *Sustainable Energy Technologies and Assessments*, vol. 56, p. 103076, Mar. 2023.

[27] B. C. Choi, H. S. Shin, S. Y. Lee, and T. Hur, “Life cycle assessment of a personal computer and its effective recycling rate,” *International Journal of Life Cycle Assessment*, vol. 11, no. 2, pp. 122–128, Mar. 2006.

[28] H. Duan, M. Eugster, R. Hischier, M. Streicher-Porte, and J. Li, “Life cycle assessment study of a Chinese desktop personal computer,” *Science of the Total Environment*, vol. 407, no. 5, pp. 1755–1764, Feb. 2009.

[29] W. Scharnhorst, H.-J. Althaus, M. Classen, O. Jolliet, and L. M. Hilty, “The end of life treatment of second generation mobile phone networks: Strategies to reduce the environmental impact,” *Environmental Impact Assessment Review*, vol. 25, no. 5 SPEC. ISS., pp. 540–566, Jul. 2005.

[30] W. Scharnhorst, L. M. Hilty, and O. Jolliet, “Life cycle assessment of second generation (2G) and third generation (3G) mobile phone networks,” *Environment International*, vol. 32, no. 5, pp. 656–675, Jul. 2006.

[31] A. Billat and D. Doeran, “Environmental Impacts of a Superconducting Quantum Computer Based on a Life Cycle Assessment,” Ecole polytechnique de Louvain, Université catholique de Louvain, 2024. Prom., 2024.

[32] ISO, “ISO 14044 : Environmental Management. Life Cycle Assessment. Requirements and Guidelines.,” *Ntc-Iso 14044*, vol. 3, no. 571. p. 47, 2006.

[33] Bluefors, “System Description, LD System,” Helsinki, FI, Version 2.0, BF1000-1234517327-10, Jan. 2022.

[34] Dell, “DELL PRECISION ™ 7920 RACK - Workstation with Intel Xeon Processor,” *Canada > Workstations > Precision Fixed Workstations > Precision 7920 Rack Workstation*, 2024. [Online]. Available: https://www.dell.com/en-ca/shop/desktop-computers/precision-7920-rack-workstation/spd/precision-7920r-workstation. [Accessed: 09-Jul-2024].

[35] Bluefors, “User Manual, LD System,” Helsinki, FI, Version 2.0, BF1000-1234517327-9, Jan. 2022.

[36] Dell, “DELL PRECISION ™ 7920 RACK - Specs Sheet.” [Online]. Available: https://www.delltechnologies.com/asset/en-us/products/workstations/technical-

support/Precision_7920_Rack_Spec_Sheet.pdf. [Accessed: 09-Jul-2024].

[37] G. Wernet, C. Bauer, B. Steubing, J. Reinhard, E. Moreno-Ruiz, and B. P. Weidema, “The ecoinvent database version 3 (part I): overview and methodology.,” *The International Journal of Life Cycle Assessment*, vol. 21, no. 9, pp. 1218–1230, Apr. 2016.

[38] A. M. Dalzell *et al.*, “Quantum algorithms: A survey of applications and end-to-end complexities,” Preprint, submitted on 4 Oct 2023, Oct. 2023.

[39] M. Lehmann, “Computer, desktop, without screen, GLO, Cut-off, U, ecoinvent database version 3.9.1.” Ecoinvent 3, St. Gallen, CH, 15-Feb-2022.

[40] Intel, “Intel Xeon Platinum 8168 Processor 33M Cache 2.70 GHz Product Specifications,” *Products Home > Product Specifications > Processors*, 2024. [Online]. Available: https://ark.intel.com/content/www/us/en/ark/products/120504/intel-xeon-platinum-8168-processor-33m-cache-2-70-ghz.html. [Accessed: 31-May-2024].

[41] Dell, “DELL PRECISION ™ 7920 RACK - Technical Guidebook.” [Online]. Available: https://www.delltechnologies.com/asset/en-us/products/workstations/technical-support/dell-precision-7920-rack-technical-guidebook.pdf. [Accessed: 31-May-2024].

[42] ecoinvent, “System Models,” *System Models > Allocation cut-off by Classification*, 14-Feb-2024. [Online]. Available: https://support.ecoinvent.org/system-models. [Accessed: 05-Jul-2024].

[43] W. Cai, Y. Ma, W. Wang, C. L. Zou, and L. Sun, “Bosonic quantum error correction codes in superconducting quantum circuits,” *Fundamental Research*, vol. 1, no. 1. KeAi Communications Co., pp. 50–67, 01-Jan-2021.

[44] P. Shi, J. Yuan, F. Yan, and H. Yu, “Multiplexed control scheme for scalable quantum information processing with superconducting qubits,” Preprint, submitted on 12 Dec 2023, Dec. 2023.

[45] TOP500, “Home | TOP500,” 2024. [Online]. Available: https://top500.org/. [Accessed: 05-Jul-2024].

[46] C. Q. Choi, “The Beating Heart of the World’s First Exascale Supercomputer,” *IEEE Spectrum*, 24-Jun-2022.

[47] A. Geist, “Update on Frontier exascale system and early science,” in *ASCAC Meeting*, 2022, p. 16.

[48] AMD, “AMD EPYC™ 7003 Series Processors,” *Advanced Micro Devices, Inc.*, 2024. [Online]. Available: https://www.amd.com/en/products/processors/server/epyc/7003-series.html. [Accessed: 05-Jul-2024].

[49] S. Scott, “The Cray Shasta architecture: Designed for the exascale era,” *2020 High Performance Computing Conference*, Youtube channel: Rice Ken Kennedy Institute, p. 27, 03-Mar-2020.

[50] C. Bulle *et al.*, “IMPACT World+: a globally regionalized life cycle impact assessment method,” *International Journal of Life Cycle Assessment*, vol. 24, no. 9, pp. 1653–1674, Sep. 2019.

[51] IPCC, *Climate Change 2021: The Physical Science Basis. Contribution of Working Group I to the Sixth Assessment Report of the Intergovernmental Panel on Climate Change (IPCC)*. Cambridge, UK and New York, NY, USA: Cambridge University Press. In Press, 2021.

[52] PRé Sustainability, “SimaPro | LCA software for informed changemakers,” 2024. [Online]. Available: https://simapro.com/. [Accessed: 01-Jul-2024].

[53] H. Neven and J. Kelly, *Suppressing quantum errors by scaling a surface code logical qubit*. 2023.

[54] ENERGY STAR, “Energy Efficient Products for Businesses,” *Energy Efficient Products*, 2024. [Online]. Available: https://www.energystar.gov/products/business. [Accessed: 26-Aug-2024].

[55] ENERGY STAR, “ENERGY STAR Certified Computers - DELL - D04X : Precision 7920 Tower,” *Energy Efficient Products - Product Finder - ENERGY STAR Certified Computers - DELL - D04X : Precision 7920 Tower*, 2017. [Online]. Available: https://www.energystar.gov/productfinder/product/certified-computers/details/2359366. [Accessed: 26-Aug-2024].

[56] B. Barber *et al.*, “A real-time, scalable, fast and highly resource efficient decoder for a quantum computer,” Preprint, [Submitted on 11 Sep 2023 (v1), last revised 24 Sep 2024 (this version, v2)], Sep. 2024.

[57] N. Liyanage, Y. Wu, S. Tagare, and L. Zhong, “FPGA-based Distributed Union-Find Decoder for Surface Codes,” in *IEEE Transactions on Quantum Engineering*, 2024, vol. 5, pp. 1–18.

[58] R. Hischier, M. Classen, M. Lehmann, and W. Scharnhorst, “Life cycle inventories of electric and electronic equipments: Production, use and disposal,” Dübendorf, CH, Ecoinvent report N° 18, 2007.

[59] L. Lannelongue, J. Grealey, and M. Inouye, “Green Algorithms: Quantifying the Carbon Footprint of Computation,” *Advanced Science*, vol. 8, no. 12, p. 2100707, Jun. 2021.

[60] Green Algorithms, “Green Algorithms Calculator,” 2024. [Online]. Available: http://calculator.green-algorithms.org/. [Accessed: 15-Aug-2024].

[61] N. Arora and P. Kumar, “Sustainable Quantum Computing: Opportunities and Challenges of Benchmarking Carbon in the Quantum Computing Lifecycle,” Preprint, Submitted on 11 Aug 2024 (v1), last revised 13 Aug 2024 (this version, v2), Aug. 2024.

[62] AMD, “AMD Instinct™ MI250X Accelerators,” *Advanced Micro Devices, Inc.*, 2024. [Online]. Available: https://www.amd.com/en/products/accelerators/instinct/mi200/mi250x.html. [Accessed: 08-Jul-2024].

[63] Nvidia, “Data Sheet: Quadro P4000,” Jun. 2018.

[64] HPE, “HPE Cray EX Supercomputer - Specifications,” 2024. [Online]. Available: https://support.hpe.com/hpesc/public/docDisplay?docId=a00109704en_us&docLocale=en_US#N10073. [Accessed: 07-Aug-2024].